\documentclass[draft]{agujournal2019}
\usepackage{url} 
\usepackage{lineno}
\usepackage[inline]{trackchanges} 
\usepackage{soul}
\usepackage{ragged2e}
\usepackage{longtable}
\usepackage{amsmath,amssymb}
\usepackage{multibib}
\draftfalse

\journalname{Manuscript submitted to Geophysical Research Letters}

\begin{document}

%
%

\title{Diffusion-free scaling in rotating spherical Rayleigh-B\'enard convection}

%
%




\authors{Guiquan Wang\affil{1}, Luca Santelli\affil{3}, Detlef Lohse\affil{1,4}, Roberto Verzicco\affil{1,2,3},\\ Richard J. A. M. Stevens\affil{1}}

\affiliation{1}{Physics of Fluids Group and Twente Max Planck Center, Department of Science and Technology, Mesa+ Institute, and J. M. Burgers Center for Fluid Dynamics, University of Twente, P.O. Box 217, 7500 AE Enschede, The Netherlands}

\affiliation{2}{Dipartimento di Ingegneria Industriale, University of Rome' Tor Vergata', Via del Politecnico 1, 00133 Rome, Italy}

\affiliation{3}{Gran Sasso Science Institute, Viale F. Crispi 7, 67100 L'Aquila, Italy}

\affiliation{4}{Max Planck Institute for Dynamics and Self-Organization, Am Fassberg 17, 37077 G\"ottingen, Germany}





\correspondingauthor{G. Wang}{gwang4academy@gmail.com}
\correspondingauthor{Richard J. A. M. Stevens}{r.j.a.m.stevens@utwente.nl}




\begin{keypoints}
\item We show that in rotating spherical Rayleigh-B\'enard convection, three regions with distinctly different flow dynamics are formed.
\item The mid-latitude region is characterized by convective columns that extend from the Northern to the Southern hemisphere of the outer sphere.
\item The diffusion-free scaling indicates that the flow dynamics and the heat transport originating in the mid-latitude region are bulk-dominated.
\end{keypoints}

\justifying
\begin{abstract}
Direct numerical simulations are employed to reveal three distinctly different flow regions in rotating spherical Rayleigh-B\'enard convection. In the \textcolor{black}{high-latitude} region $\mathrm{I}$ vertical (parallel to the axis of rotation) convective columns are generated between the hot inner and the cold outer sphere. The mid-latitude region $\mathrm{II}$ is dominated by vertically aligned convective columns formed between the Northern and Southern hemispheres of the outer sphere. The diffusion-free scaling, which indicates bulk-dominated convection, originates from this mid-latitude region. In the equator region $\mathrm{III}$ the vortices are affected by the outer spherical boundary and are much shorter than in region $\mathrm{II}$.
\end{abstract}

\section*{Plain Language Summary}
Thermally driven turbulence with background rotation in spherical Rayleigh-B\'enard convection is found to be characterized by three distinctly different flow regions. The diffusion-free scaling, which indicates the heat transfer is bulk-dominated, originates from the mid-latitude region in which vertically aligned vortices are stretched between the Northern and Southern hemispheres of the outer sphere.\ These results show that the flow physics in rotating convection are qualitatively different in planar and spherical geometries. This finding underlines that it is crucial to study convection in spherical geometries to better understand geophysical and astrophysical flow phenomena.

\section{Introduction}

\begin{figure}
\centering
\includegraphics[width=13cm, trim={0cm 0cm 0cm 0cm},clip]{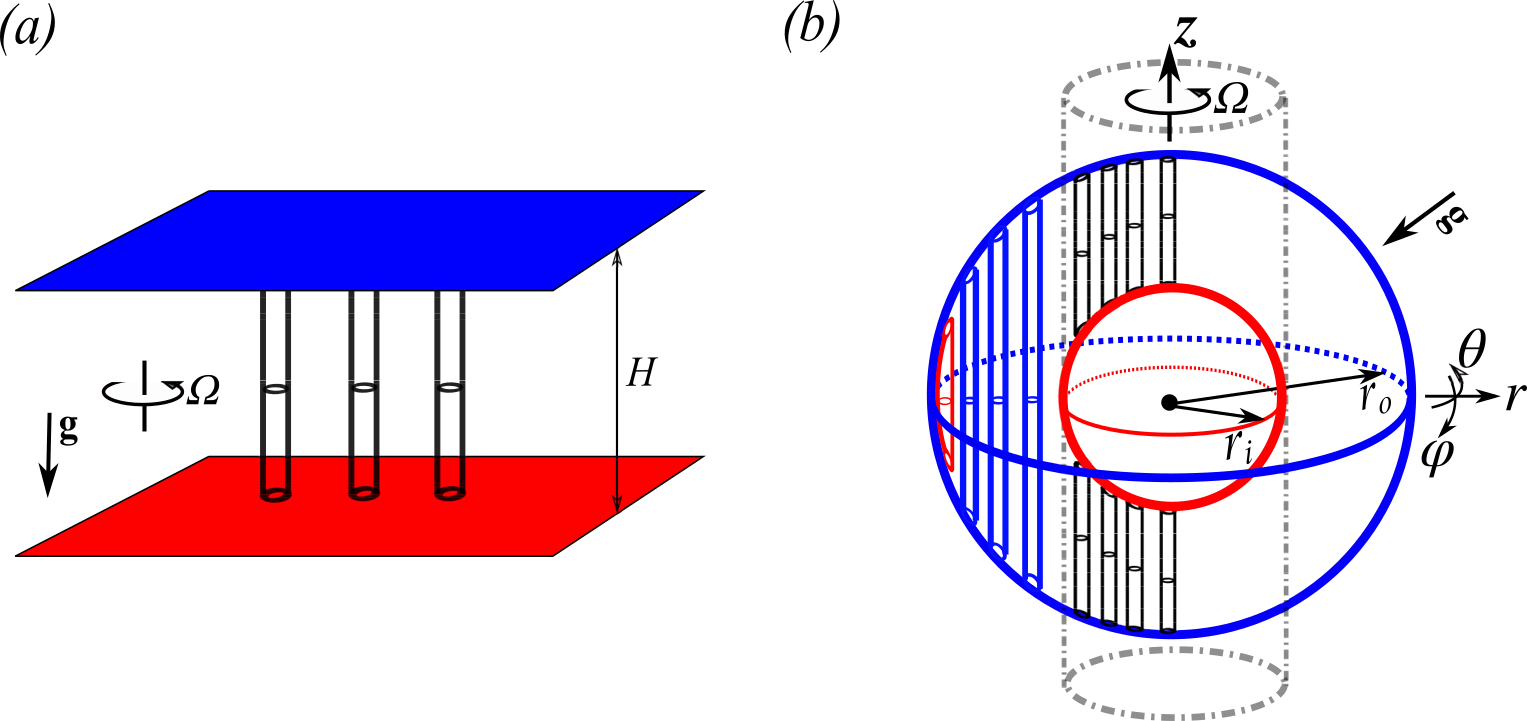}
\caption{Schematics showing the alignment of the axial convective columns in (a) \textcolor{black}{canonical framework heated from bottom and cooled from above} and (b) spherical rotating RB convection heated from inner and cooled from outer, in which the gravity points towards the geometrical centre. The \textcolor{black}{longitudinal (azimuthal), co-latitudinal (polar)}, and radial directions are represented by $\hat{\theta}$, $\hat{\varphi}$ and $\hat{r}$, respectively. The rotation axis aligns with the $z$-direction. The angle between gravity and rotation axis is $\varphi$. The tangent cylinder is shown with dashed-dotted grey line. Panel (b) is adapted from \citeA{busse1970thermal, busse1983model, aurnou2015rotating}.}
\label{fig:planar_vs_spherical}
\end{figure}

{\color{black}
Rapidly rotating convection is relevant for many geophysical and astrophysical flows, e.g.\ the solar interior \cite{sch20}, the liquid metal core of terrestrial planets \cite{zhang2000magnetohydrodynamics,olson2011laboratory, jones2011planetary, aurnou2015rotating}, and Earth's oceans and atmosphere \cite{marshall1999open,fultz1959studies}. In these instances of convection with strong thermal driving, the flow dynamics is nevertheless dominated by the strong background rotation \cite[]{sprague2006numerical,aurnou2015rotating,kunnen2021geostrophic}. The effect of rotation has been extensively studied in Rayleigh-B\'enard (RB) convection experiments \cite{rossby1969study, liu1997heat, stevens2009transitions, king2009boundary, king2012heat, zhong2009prandtl, ecke2014heat, stellmach2014approaching, Cheng2020,wed21} and simulations \cite{king2009boundary, king2012heat, king2013scaling, schmitz2009heat, stevens2009transitions, stellmach2014approaching, horn2015toroidal, kunnen2016transition}. In the canonical RB system, the flow is confined between two parallel plates, and this system is studied in 3D periodic, rectangular, or cylindrical domains. In the remainder of this paper, we refer to this as planar RB convection to distinguish it from the spherical RB system considered in here (see figure \ref{fig:planar_vs_spherical}(a)). We refer the reader to the reviews \cite{aurnou2015rotating, plumley2019scaling,kunnen2021geostrophic} for an extensive explanation of rotating RB convection. Even though there are great community efforts on rotating RB the diffusion-free scaling regime, geostrophic dominated
which will be defined explicitly below, predicted for strongly thermally driven rotation dominated flow has not been observed yet for rotating RB with no-slip boundaries. This study will show that in a spherical RB convection, the geometry allows for the formation of a geostrophic dominated flow region that exhibits diffusion-free scaling in the mid-latitude region.
}

The control parameters of rotating RB flow are the Rayleigh ($Ra$), Ekman ($Ek$), and Prandtl ($Pr$) numbers, to be defined explicitly below. Derived from these, the convective Rossby number $Ro \equiv \sqrt{{Ra}/{Pr}} {Ek}/{2}$ characterizes the importance of the thermal forcing relative to rotation \cite{gilman1977nonlinear}. With increasing Rayleigh number $Ra$ and for strong rotation $Ro\ll1$, two regimes can be identified, namely: (1) the \emph{weakly nonlinear regime} for $Ra$ near the onset of convection, (2) the \emph{quasi-geostrophic regime} for $Ra/Ra_c \leq 3$ \cite{ecke2014heat}, where $Ra_c \sim Ek^{-4/3}$ is the critical Rayleigh number for the onset of convection \cite{CHANDRASEKHAR1961}. In a third regime (3), for $Ro \gg 1$ and high enough $Ra$, the flow approaches the non-rotating RB convection case  \cite{grossmann2000scaling, RevModPhys.81.503, Chilla2012}. 

For the quasi-geostrophic regime, when $Ek\rightarrow0$, {\color{black} the Nusselt number $Nu$ (i.e.\ the non-dimensional heat transfer) is found to depend on the supercriticality $Nu\sim Pr^\gamma (Ra/Ra_c)^{\alpha}$ \cite{king2012heat, julien2012heat, stellmach2014approaching, cheng2015laboratory}. When the heat transport is independent of molecular diffusion in the asymptotic limit, this results in $\alpha=3/2$ and $\gamma=-1/2$. This scaling $Nu\sim Pr ^{-1/2}(Ra/Ra_c)^{3/2}$ is known as \emph{diffusion-free scaling}. The physics of the diffusion-free scaling, similar to the ultimate regime in RB convection \cite{kraichnan1962turbulent, spiegel1971convection, shraiman1990heat, grossmann2011multiple}, is that the thermal and kinetic boundary layers, and thus the kinematic viscosity and thermal diffusivity, do not play an explicit role anymore for the heat flux scaling. This is known as bulk-dominated convection.} 

So far, the diffusion-free scaling has only been obtained in planar convection by considering an asymptotically reduced model in which Ekman pumping effects are not represented \cite{julien2012heat} and numerical simulation with free-stress boundaries and $Ek \le 10^{-6}$ \cite{stellmach2014approaching, kunnen2016transition}. For planar convection with no-slip boundaries, \citeA{king2012heat,king2013scaling} theoretically predict $\alpha=3$ for $Ra\lesssim Ek^{-3/2}$. This finding follows from an analysis of the boundary layer stability and is supported by experimental and simulation data for $10^{-6} \le Ek \le \infty$. The difference between $\alpha=3$ for no-slip boundaries and $\alpha=3/2$ for free-stress boundaries is attributed to the active role of the Ekman pumping in the boundary layers near the plates \cite{plumley2016effects, julien2016nonlinear}. However, the asymptotic diffusion-free scaling exponent $\alpha=3/2$ has not been reported for no-slip boundaries in planar convection. 

However, \citeA{Gastine2016} find the diffusion-free scaling for $Ek \leq 10^{-5}$ for $6Ra_c \leq Ra \leq 0.4Ek^{-8/5}$ in spherical RB convection with inner-to-outer radius ratio $\eta=0.6$ and no-slip boundaries. The $Ek^{-8/5}$ scaling is proposed by \citeA{julien2012heat,julien2012statistical}. We note that previous theories of \citeA{gilman1977nonlinear} (giving the transitional Rayleigh number $Ra_t \sim Ek^{-2}$ where $Ra_t$ represents for the upper bound of the diffusion-free scaling region) and of \citeA{king2009boundary} (giving $Ra_t \sim Ek^{-7/4}$) do not appropriately capture the upper bound of the diffusion-free scaling region, which scales as $Ek^{-8/5}$.

The objective of this work is to elucidate the observation of diffusion-free scaling in spherical RB convection at relatively weak rotation ($Ek \sim 10^{-5}$), while this scaling is not observed in planar convection. For strong rotation $Ro\ll1$, the Taylor Proudman effect \cite{taylor1923viii} favors invariance along the rotation axis. In planar convection, see figure \ref{fig:planar_vs_spherical}(a), the rotation axis is orthogonal to the plates, and the convective columns are homogeneously distributed in the horizontal direction and always stretch between the hot and cold plates. However, in spherical geometry, the rotation effect is latitude dependent, see figure \ref{fig:planar_vs_spherical}(b), due to which three distinctly different flow regions are formed. Inside the inner sphere's tangent cylinder, the convective columns touch the inner and outer spherical boundaries. In the mid-latitude region the convective columns are stretched between the Northern and Southern hemispheres of the outer sphere. Near the equator, the convective columns adjust themselves to the curved boundary. This work will show that the diffusion-free scaling originates from this mid-latitude region. The paper is organized as follows: \textcolor{black}{In Section \ref{sec:Numerical}, we introduce the rotating spherical RB system with its control parameters. Section \ref{sec:Nu_overview} is an overview of our simulation results compared and validated to literature, subsequent analysis is performed in Sections \ref{sec:Three _regions} and \ref{sec:Region_II}. Finally, we conclude our findings in Section \ref{sec:Conclusions}.}

\section{Numerical method, control and response parameters}\label{sec:Numerical}

A sketch of the rotating spherical RB geometry is shown in figure \ref{fig:planar_vs_spherical}(b). A fluid fills a spherical shell between the inner sphere of radius $r_i$ and outer sphere of radius $r_o$ with distance $d=r_o-r_i$ from the inner one. The whole system rotates about the vertical $z$ axis at angular velocity $\Omega$. The surface temperature of the inner and outer spheres is kept constant at $T_i$, and $T_o$, respectively, with $T_i > T_o$. No-slip boundary conditions are imposed at both spheres. We solve the Navier-Stokes equations in spherical coordinates within the Boussinesq approximation, which in dimensionless form read:
\begin{equation}
 \frac{\partial \mathbf{u}}{\partial t}+\mathbf{u} \cdot \nabla \mathbf{u} =-\nabla p+\sqrt{\frac{P r}{R a}} \nabla^{2} \mathbf{u}+g T \vec{\mathbf{e}}_r-\frac{1}{Ro}\vec{\mathbf{e}}_z \times \mathbf{u} ~~,~~\nabla \cdot \mathbf{u}=0, \\
 \label{eq:NS_2}
\end{equation}
\begin{equation}
 \frac{\partial T}{\partial t}+\mathbf{u} \cdot \nabla T=\frac{1}{\sqrt{{Ra Pr}}} \nabla^{2} T.
 \label{eq:NS_3}
\end{equation}
where $\mathbf{u}(\vec{x},t)$, $p(\vec{x},t)$, $T(\vec{x},t)$, and $g(r)$ denote the fluid velocity, pressure, temperature and radially dependent gravitational acceleration. 

\textcolor{black}{In this study we focus on a radius ratio $\eta=r_i/r_o=0.6$ and the gravity profile $g(r) \sim (r_o/r)^2$ valid for homogeneous mass distribution to allow comparisons with non-rotating \cite{Gastine2015} and rotating \cite{Gastine2016} convection in spherical RB. This system configuration is considered representative for studying convection in gas giants \cite{long_mound_davies_tobias_2020}. Additionally, we perform simulations for $\eta=0.35$ and $g(r) \sim (r_o/r)^{-1}$, which is considered an Earth-like configuration used by \citeA{yadav2016effect} and \citeA{long_mound_davies_tobias_2020}.} The equations are discretized by a staggered central second-order finite-difference scheme in spherical coordinates \cite{santelli2020finite}. We use a uniform grid in the longitudinal and co-latitudinal directions and ensure that the bulk and boundary layers are appropriately resolved \cite{ste10}. The grid cells are clustered towards the inner and outer sphere to ensure the boundary layers are adequately resolved \cite{shishkina2010boundary}. Further details on the simulations are given in the supplementary material.

The dynamics of rotating spherical RB convection are determined by the Rayleigh, Prandtl, and Ekman numbers
\begin{equation}
Ra=\frac{\beta g_o d^3 \Delta T}{\kappa \nu},~ Pr=\frac{\nu}{\kappa},~Ek=\frac{\nu}{\Omega d^2},
\label{eq:Ra_Nu_E_Ro}
\end{equation}
where $\beta$ is the thermal expansion coefficient, $g_o$ is the gravity at the outer sphere, $\nu$ is the kinematic viscosity, and $\kappa$ is the thermal diffusivity of the fluid. $Ra$ is a measure of the thermal driving of the system, $Ek$ characterizes the ratio of viscous to Coriolis forces, and $Pr$ indicates the ratio of the viscous to thermal diffusivities. In this study we consider $Pr=1$. We use the Rossby number $Ro \equiv \sqrt{{Ra}/{Pr}} {Ek}/{2}$ to evaluate the relative importance of rotation and buoyancy \cite{gilman1977nonlinear}. We normalize the results using the length scale $d=r_o-r_i$, the temperature difference $\Delta T$ between inner and outer sphere, and the free-fall velocity $U=\sqrt{\beta g_o \Delta T d}$.

The Nusselt number quantifies the non-dimensional heat transport
\begin{equation}
  Nu=\frac{\overline{\langle u_rT \rangle_s}-\kappa \partial_r \overline{\langle T \rangle_s}}{-\kappa \partial_r T_c},
  \label{eq:Nu_h}
\end{equation}
{\color{black}where $T_c(r)=\eta/[(1-\eta)^2r]-\eta/(1-\eta)$ is the conductive temperature profile in spherical shells with constant temperature boundary conditions $T_c(r_i)=1$ and $T_c(r_o)=0$. The notations $\langle \dotsb \rangle_s$ represents the average over a spherical surface with constant distance from the center, e.g. ${\langle T \rangle_s}=\frac{1}{4\pi} {\int_0^{2\pi}} {\int_0^\pi} T(\theta,r,\varphi) \sin{\varphi} \mathrm{d} \varphi \mathrm{d} \theta $. Overbar $\overline{\dotsb}$ corresponds to time averaging}. In the following discussion, we will use $Nu$ on the outer sphere as a function of the co-latitude

\begin{equation}
  Nu(\varphi)=\left. -\frac{1}{\eta} \frac{d \overline{\langle T \rangle_\theta}}{dr} \right| _{r_o}
  \label{eq:Nu_o}
\end{equation}
where $\langle ... \rangle_\theta$ represents the average over the azimuthal direction, e.g. ${\langle T \rangle_\theta}=\frac{1}{2\pi} {\int_0^{2\pi}} T(\theta,r,\varphi) \mathrm{d} \theta $.

\section{Heat transfer in rotating spherical RB convection}\label{sec:Nu_overview}

\begin{figure}
\centering
\includegraphics[width=14cm, trim={0cm 0cm 0cm 0cm},clip]{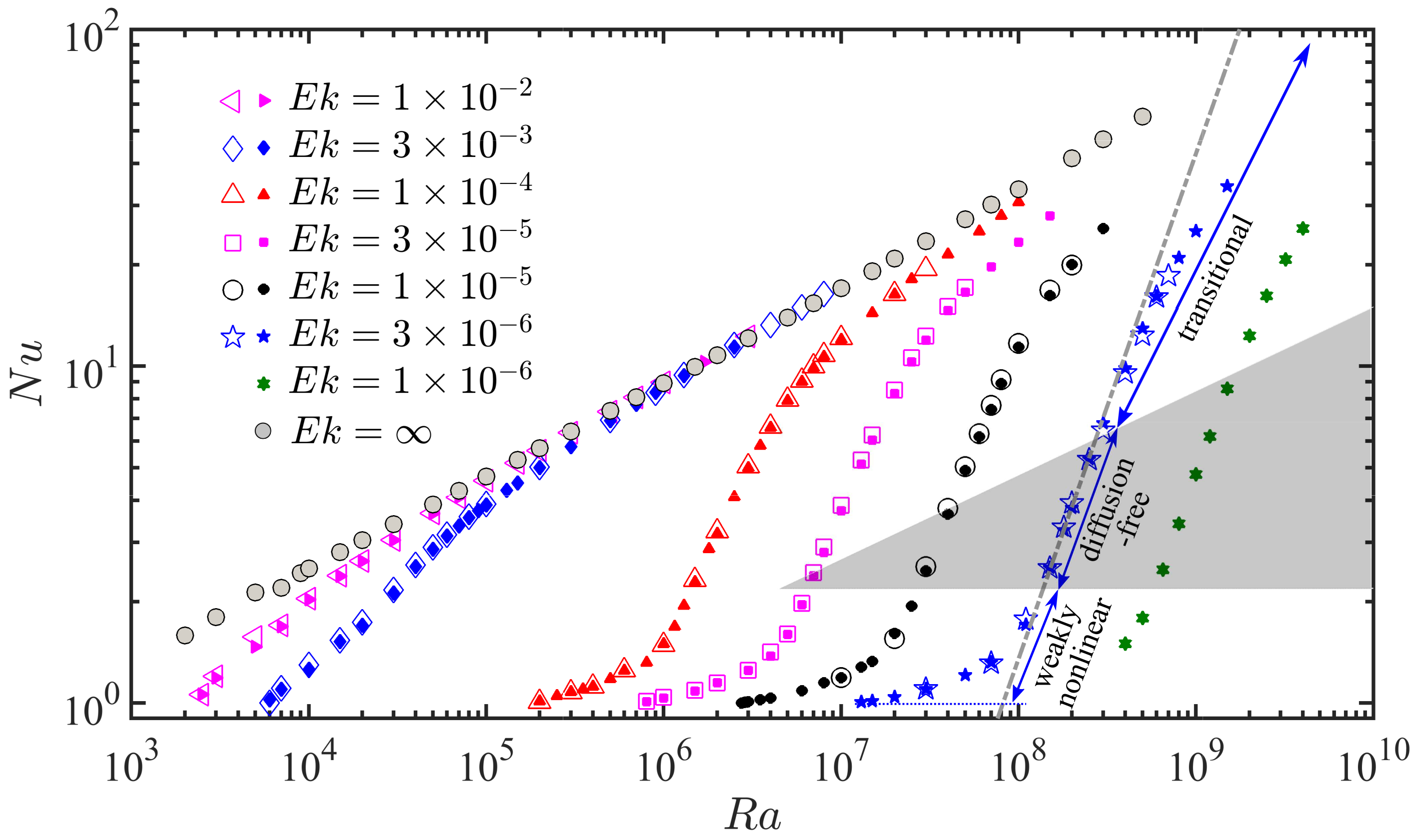}
\caption{$Nu$ as function of $Ra$ for different $Ek$. Rotating cases: open symbols indicate the present results, filled-in symbols are those from \citeA{Gastine2016}.\ Non-rotating cases from \citeA{Gastine2015} are indicated by $Ek=\infty$. The shaded wedge-shaped region indicates the diffusion-free scaling regime ($6Ek^{4/3} \leq Ra \leq 0.4Ek^{-8/5}$), which corresponds to the quasi-geostrophic regime identified by \citeA{Gastine2016}. The dot-dashed grey line gives the diffusion-free scaling $Nu=0.149R^{3/2}$ for $Ek=3\times10^{-6}$. The error bars are smaller than the symbol sizes.}
\label{fig:Nu_overview}
\end{figure}

Figure \ref{fig:Nu_overview} shows $Nu$ as function of $Ra$ for various $Ek$. The results from our simulations agree excellently with those from \citeA{Gastine2016}. For strong enough rotation (e.g.\ $Ek \leq 3\times 10^{-5}$), with increasing $Ra$ three regimes can be identified \cite{Gastine2016, long_mound_davies_tobias_2020}. For low $Ra$, in the weakly nonlinear regime, rotational effects are dominant ($Ro\ll1$) and $Nu \sim R ^ \alpha$ with $R \equiv Ra Ek^{4/3}$ and $\alpha=1$. In the quasi-geostropic regime with diffusion-free scaling $\alpha=3/2$, the Taylor-Proudman effect favours invariance along the rotation axis, thereby suppressing global heat transport relative to non-rotating case \cite{julien2012heat}. This regime is observed for $6Ek^{4/3} \leq Ra \leq 0.4Ek^{-8/5}$ \cite{Gastine2016}. The lower bound is related to $Ra_c$, while the upper bound corresponds to the asymptotic prediction for bulk-limited heat transfer in geostrophic turbulence by \citeA{julien2012heat}. In the transitional regime between strong and weak rotation ($Ro \sim 1$) the buoyancy force gradually becomes dominant over rotational effects with increasing $Ra$ and the flow eventually approaches the non-rotating case for $Ro\gg1$.

\section{Identification of three flow regimes}\label{sec:Three _regions}

\begin{figure}
\centering
\includegraphics[width=14.1cm, trim={0cm 0cm 0cm 0cm},clip]{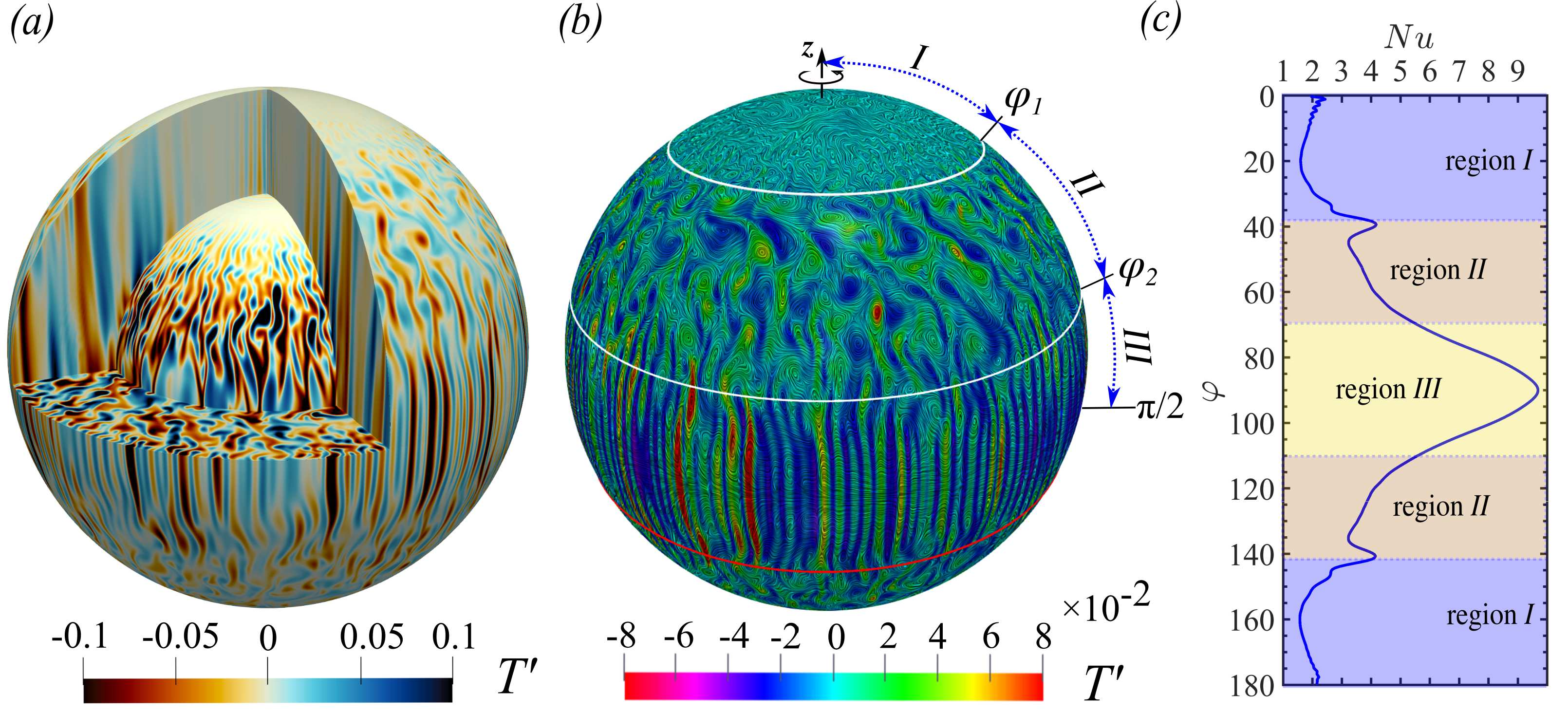}
\caption{(a) Contour of the temperature fluctuation $T'$ on two meridional cuts, equatorial section, and two spherical surfaces (\textcolor{black}{corresponding to the spherical surfaces located at the inner ($r=r_i+\lambda_{T,i}$) and outer ($r=r_o-\lambda_{T,o}$) thermal boundary layers}).\ (b) Contour of $T'$ with streamlines illustrated by using line integral convolution on the outer radial surface \textcolor{black}{(see Section 3 in the supplementary material).} The definition of the three regimes $\mathrm{I, II, III}$ is given in the text and figure \ref{fig:Zonal_Flow}.\ (c) $Nu$ as function of the co-latitude $\varphi$ on the outer sphere. In all cases (a)-(c), $Ek=1\times10^{-5}$ and $Ra=5\times10^{7}$, i.e.\ simulation No.76 in the supplementary material.}
\label{fig:contour_3_Regions}
\end{figure}

\textcolor{black}{Figure \ref{fig:contour_3_Regions}(a) visualizes the columnar structures by $T'>0$ and $T'<0$, here $T'(\theta,r,\varphi)=T(\theta,r,\varphi)-\overline{\langle T \rangle_s}$, $\overline{\langle T \rangle_s}$ is defined in Section 2. The inner and outer thermal boundary layer thickness $\lambda_{T,i}$ and $\lambda_{T,o}$ is defined by the intersection of the linear fit to ${\langle T \rangle_s}$ near the boundaries and the profile at mid-depth \cite{Gastine2016, long_mound_davies_tobias_2020}}. Figure \ref{fig:contour_3_Regions}(b) clearly shows that there are three distinct flow regions. Region $\mathrm{I}$ spans from the rotation axis to $\varphi_1$, where $\varphi_1$ can be determined by the intersection between the cylinder tangent to the inner sphere with the outer sphere. In this region, the columnar structures connect the boundary layers around the inner and outer spheres. Region $\mathrm{II}$ is found between $\varphi_1$ and $\varphi_2$ (see figure \ref{fig:contour_3_Regions}(b)), $\varphi_2$ being the maximum zonal flow location (see below). In this mid-latitude region, the structures are the strongest, and tall thin columns stretch from the Northern to the Southern parts of the cold outer sphere. Region $\mathrm{III}$ is the region around the equator, see figure \ref{fig:contour_3_Regions}(b). In this region, the structures aligned with the rotation axis are much shorter than in the mid-latitude region $\mathrm{II}$, while they conform themselves to the outer spherical boundary. Figure \ref{fig:contour_3_Regions}(c) shows that the heat transport strongly depends on the latitude \cite{yadav2016effect}, which means that the heat transfer in the different flow regions identified above is different.

\begin{figure}
\centering
\includegraphics[width=14cm, trim={0cm 0cm 0cm 0cm},clip]{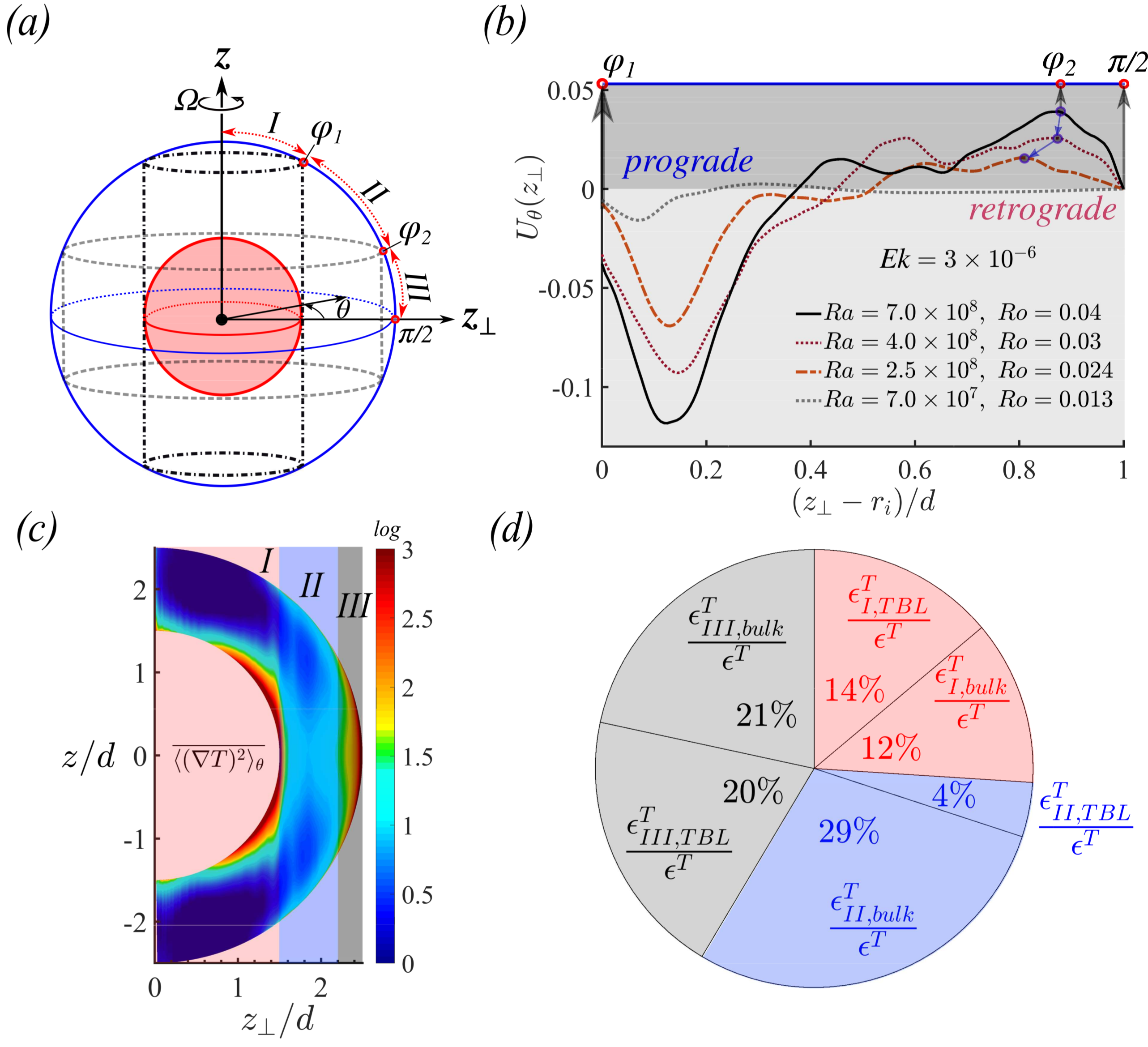}
\caption{ (a) Cylindrical polar coordinates $(z,z_\perp,\theta)$, $z$ is the rotation axis, $z_ \perp$ is the cylindrical radius and $\theta$ is the azimuthal angle and of which the regimes $\mathrm{I, II, III}$ can be defined as shown. 
(b) Ensemble averaged azimuthal velocities $U_\theta$ (zonal flows) as a function of $z_ \perp$ in equation (\ref{eq:Zonal_flow}). $(z_\perp-r_i)/d=0$ and $1$ correspond to the tangent cylinders of the inner and outer spheres, respectively. $\varphi_2$ is determined by the $z_\perp$ location close to the outer sphere ($(z_\perp-r_i)/d=1$) where the zonal flow is strongest. (c) Time and azimuthal averaged thermal dissipation $\overline{\langle (\nabla T)^2 \rangle_{\theta}}$ in the meridional plane for case No.76 of $Ek=1\times10^{-5}$ and $Ra=5\times10^7$. 
(d) Pie chart for (c) showing the distribution of the thermal dissipation rate over the different regions in the boundary layer and bulk, see equation (\ref{eq:diss_2}).}
\label{fig:Zonal_Flow}
\end{figure}

{\color{black}
\citeA{aurnou2001strong} and \citeA{christensen_2002} found that the zonal flow is prograde in the equatorial region near the outer boundary and retrogrades near the tangent cylinder that encloses the central core. Therefore, the zonal flow is suitable to identify the boundary between region $\mathrm{II}$ and $\mathrm{III}$.} Figure \ref{fig:Zonal_Flow} (a,b) show how we use the local maximum prograde zonal velocity close to the equator to set $\varphi_2$. Figure \ref{fig:Zonal_Flow}(a) illustrates the cylindrical coordinate system $(z,z_\perp,\theta)$ that is used to represent the zonal flow in figure \ref{fig:Zonal_Flow}(b). The zonal flow is the ensemble average of the azimuthal velocity in cylindrical coordinate
\begin{equation}
U_\theta(z_\perp)=\overline{\langle u_\theta(z,z_\perp,\theta)\rangle_{\theta,z}}
\label{eq:Zonal_flow}
\end{equation}
where $u_\theta(z,z_\perp,\theta)$ is the longitudinal velocity $u_\theta(\theta,r,\phi)$ in spherical coordinate projected to cylindrical coordinate, $\langle \dotsb \rangle_{\theta,z}$ indicates spatial average over a cylindrical surface (in the azimuthal and vertical direction), and $\overline{\dotsb}$ indicates time-averaging. 

{\color{black}
We analyze the thermal dissipation in the different flow regions to determine whether the different regions are dominated by the boundary layer or the bulk dynamics. For spherical shells with radius ratio $\eta$, the thermal dissipation rate
\begin{equation}
\epsilon^T \equiv \overline{\langle (\nabla T)^2 \rangle} = \dfrac{3\eta}{1+\eta+\eta^2} Nu
\label{eq:diss_1}
\end{equation}
by volume integral of $T\times$(\ref{eq:NS_3}). Figure \ref{fig:Zonal_Flow}(c) shows the time-averaged thermal dissipation rate in the meridional plane. The figure shows that the thermal dissipation intensity is highest in the boundary layers along the inner sphere (region $\mathrm{I}$) and close to the equator region along the outer sphere (region $\mathrm{III}$). We determine the distribution of the thermal dissipation rate over the different regions as follows
\begin{equation}
\epsilon^T=\epsilon^T_{I,bulk}+\epsilon^T_{I,TBL}+\epsilon^T_{II,bulk}+\epsilon^T_{II,TBL}+\epsilon^T_{III,bulk}+\epsilon^T_{III,TBL,}
\label{eq:diss_2}
\end{equation}
where bulk indicates the bulk regions and TBL indicates the thermal boundary layer regions, i.e. for the radial locations r; $ r_i \leq r \leq r_i+\lambda_{T,i}$ along the inner sphere and $r_o-\lambda_{T,o} \leq r \leq r_o$ along the outer sphere. Figure \ref{fig:Zonal_Flow}(d) confirms that regions I and III are both strongly affected by the boundary layer dynamics. However, region $\mathrm{II}$ turns out to be bulk-dominated. We note that the boundary between region $\mathrm{II}$ and $\mathrm{III}$ is not determined based on the thermal dissipation profiles as there is not a clear peak in the direction separating the regimes. Therefore, as discussed above, we use the maximum in the zonal flow profile to determine this transition.

In the following section, we will show that, in agreement with theoretical expectations discussed above, the scaling of the heat transfer in the region $\mathrm{II}$ follows the diffusion-free scaling for rotation dominated strongly thermally driven flows.
}
\section{Diffusion-free scaling in region $\mathrm{II}$}\label{sec:Region_II}

\begin{figure}[htb]
\centering
\includegraphics[width=14.2cm, trim={0cm 0cm 0cm 0cm},clip]{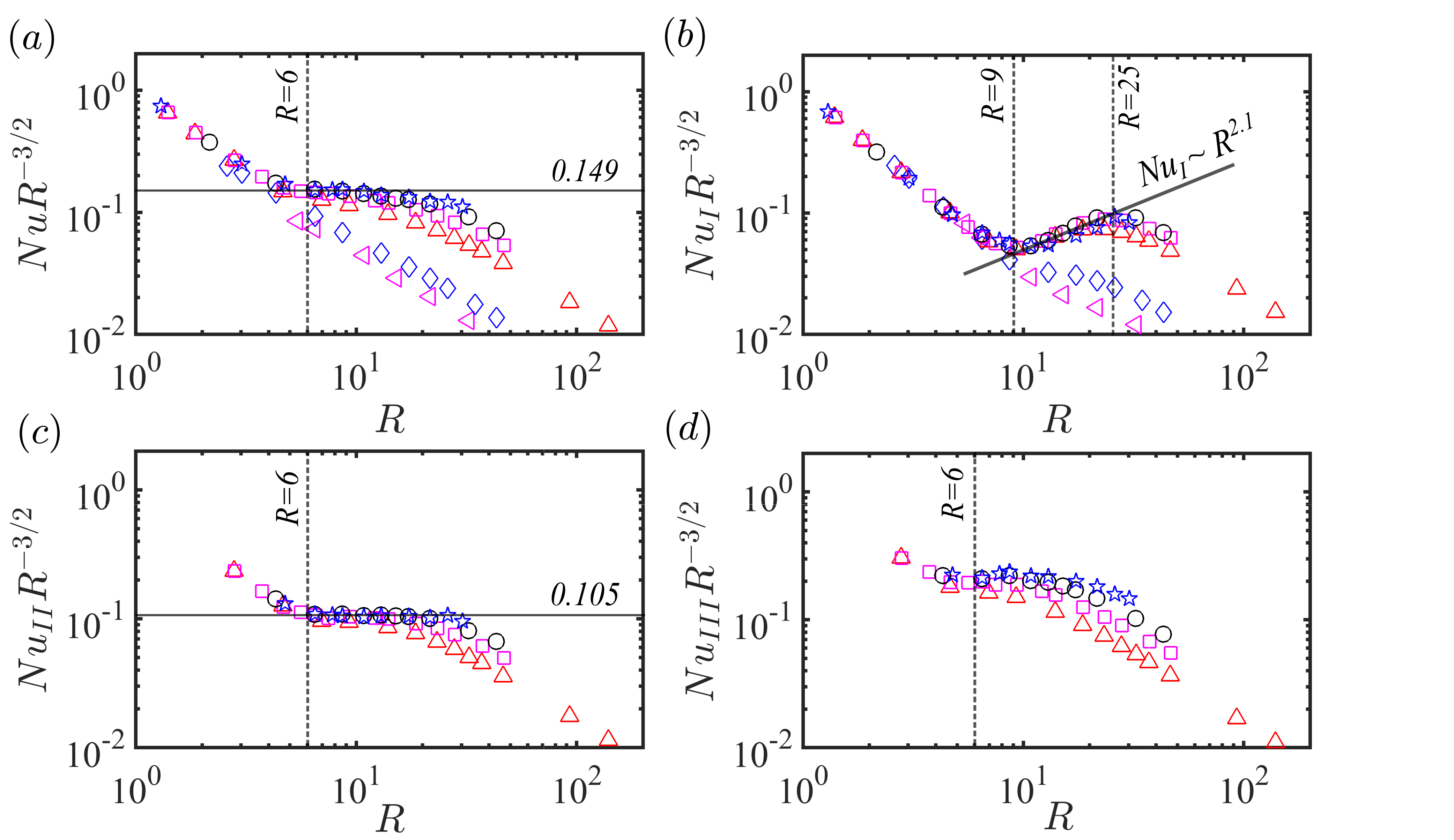}
\caption{$Nu$ on the outer sphere compensated by $R^{-3/2}$ and as a function of $R \equiv RaEk^{4/3}$.\ (a) Integration over the whole sphere; (b-d) $Nu$ in regions $\mathrm{(I-III)}$, see figure \ref{fig:contour_3_Regions}(b).\ The symbols have the same meaning as in figure \ref{fig:Nu_overview}.}
\label{fig:Nu_Regions}
\end{figure}

\begin{figure}[htb]
\centering
\includegraphics[width=10cm, trim={0cm 0cm 0cm 0cm},clip]{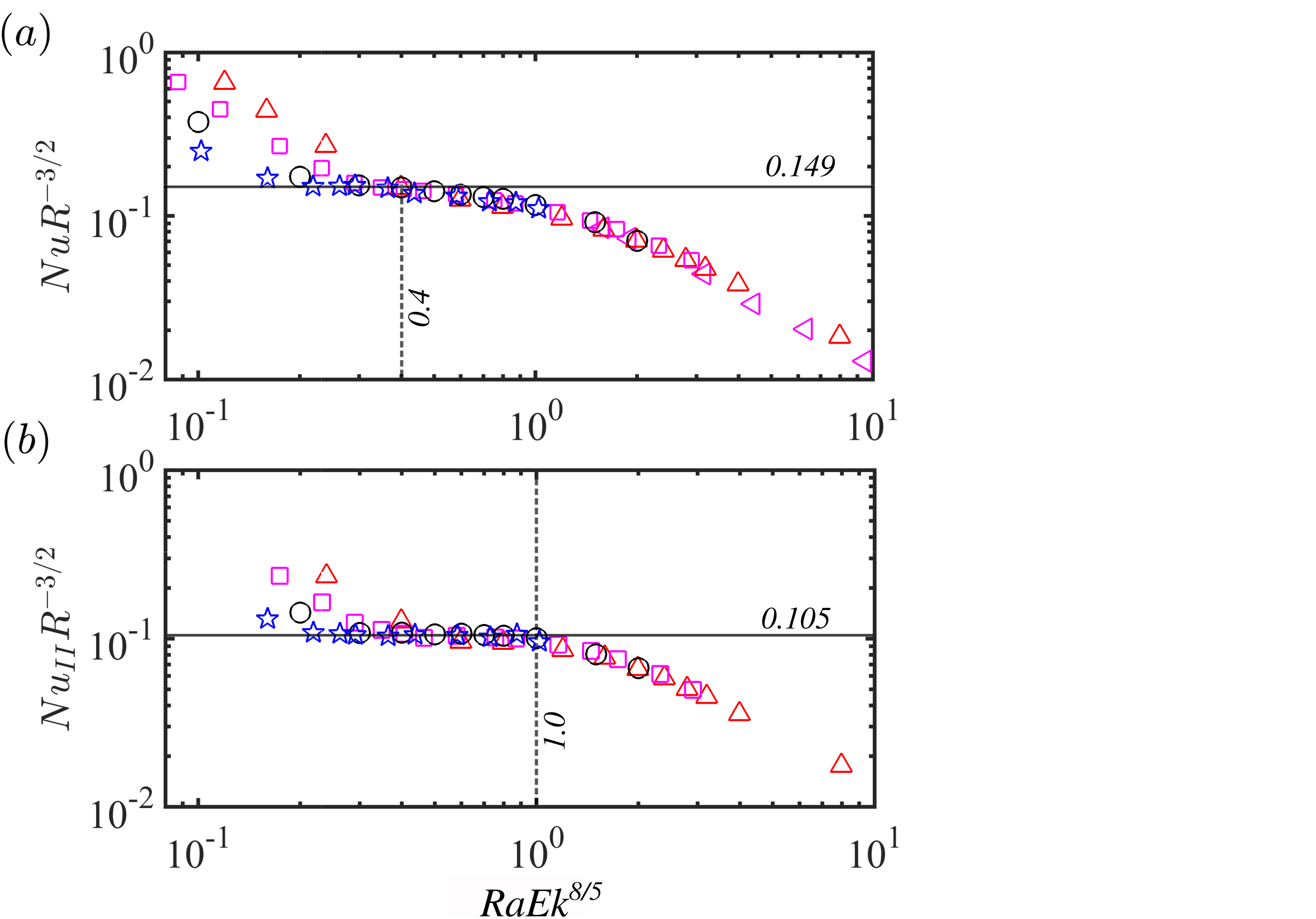}
\caption{$Nu$ compensated by $R^{-3/2}$ as a function of $Ra Ek^{8/5}$. (a) Integration over the whole sphere. The horizontal line is $Nu R^{-3/2}=0.149$ and the vertical line is $Ra Ek^{8/5}=0.4$; (b) Region $\mathrm{II}$. The horizontal line is $Nu R^{-3/2}=0.105$ and the vertical line is $Ra Ek^{8/5}=1$.\ The symbols have the same meaning as in figure \ref{fig:Nu_overview}.}
\label{fig:upper_bound}
\end{figure}

Figure \ref{fig:Nu_Regions} shows $Nu$ on the outer sphere compensated with the diffusion-free scaling law. Panel \ref{fig:Nu_Regions}(a) shows that for the global heat transfer and $Ek \leq 5 \times 10^{-5}$ the diffusion-free scaling is observed for $R \geq 6$. The crossover from the quasi-geostrophic region to the transitional region is observed at $Ra_t=0.4Ek^{-8/5}$ \cite{Gastine2016}. Figures \ref{fig:Nu_Regions}(b-d) show the heat transfer scaling in the different flow regions identified above. Panel \ref{fig:Nu_Regions}(b) evidences that, due to Ekman pumping \cite{stellmach2014approaching, zhong2009prandtl, stevens2010optimal,stevens2013heat}, the heat transport scaling in region $\mathrm{I}$ is $Nu_\mathrm{I} \sim R^{2.1}$. This is steeper than the $\alpha=3/2$ scaling for diffusion-free convection, but shallower than the $\alpha=3$ value observed in planar convection \cite{king2013scaling}. Most importantly, panel \ref{fig:Nu_Regions}(c) shows that the diffusion-free scaling is much more pronounced in region $\mathrm{II}$ than in region $\mathrm{I}$. Although the diffusion-free scaling still starts at $R=6$, it continues for much higher $R$ than the global heat transfer, see figure \ref{fig:Nu_Regions}(a). Panel \ref{fig:Nu_Regions}(d) shows that no diffusion-free scaling regime is observed in region $\mathrm{III}$.

The diffusion-free scaling regime is observed from $6R$ up to $Ra_t$, where $Ra_t$ indicates the $Ra$ number at which the regime for bulk-limited heat transfer in geostrophic turbulence ends \cite{julien2012heat,julien2012statistical}. It was demonstrated \cite{Gastine2016} that for the global heat transfer the diffusion-free scaling regime is observed up to $Ra_t =0.4 Ek^{-8/5}$, see also figure \ref{fig:upper_bound}(a). For region $\mathrm{II}$, figure \ref{fig:upper_bound}(b) shows that the diffusion-free scaling is observed up to $Ra_t=Ek^{-8/5}$, which is considerably higher $Ra$ than for the global heat transport. 

\textcolor{black}{In the Section 4 of the supplementary material, we show that the observation of the diffusion-free scaling in the mid-latitude region $\mathrm{II}$ does not depend on the specific $\eta=0.6$, $g(r)\sim(r_o/r)^{2}$ considered here. The same conclusion is obtained by analyzing $\eta=0.35$, $g(r)\sim(r_o/r)^{-1}$ and $Ek=1\times10^{-5}$.}

\section{Conclusions}\label{sec:Conclusions}

In conclusion, we have shown that rotating spherical RB convection has three distinctly different flow regions; see figure \ref{fig:contour_3_Regions}(b). In region $\mathrm{I}$, convective columns are formed between the hot inner and cold outer spheres. The mid-latitude region $\mathrm{II}$ is the region where the vertically aligned vortices are strongest, and the flow is bulk dominated. Region $\mathrm{III}$ is formed around the equator, and here the vortices are shorter and are affected by the outer spherical boundary.
 
The diffusion-free scaling $Nu \sim (RaEk^{4/3})^{\alpha}$ with $\alpha=3/2$ originates from the mid-latitude flow region in which the flow dynamics are bulk dominated. In this region, thin and long convective columns are formed between the Northern and Southern parts of the cold outer sphere. This geostrophically dominated flow region can be formed due to the system geometry. Due to the curvature effects in spherical geometries, the latitude-dependent Coriolis force results in inhomogeneous convective columns in the co-latitudinal direction and more convective columns on the outer sphere than the inner sphere.

\acknowledgments
\textcolor{black}{The authors thank the two anonymous referees for constructive comments that improved the manuscript.} G.W.\ thanks Dr.\ Kai Leong Chong and Dr.\ Chong Shen Ng for insightful discussions. G.W. and R.J.A.M.S. acknowledge the financial support from ERC (the European Research Council) Starting Grant No.\ 804283 UltimateRB.\ This work was sponsored by NWO Science for the use of supercomputer facilities. We also acknowledge the national e-infrastructure of SURFsara, a subsidiary of SURF cooperation, the collaborative ICT organization for Dutch education and research, and Irene at Tr{\`e}s Grand Centre de Calcul du CEA (TGCC) under PRACE project 2019215098. 

\section*{Open Research}

\noindent Data Availability Statement

\noindent The data used in this paper are available for download at

\noindent https://doi.org/10.5281/zenodo.5034407

\bibliography{./agusample.bib}







\end{document}